\documentclass[runningheads]{llncs}
\usepackage[T1]{fontenc}
\usepackage{graphicx}
\usepackage{amsmath}
\usepackage{hyperref}
\begin{document}
\title{DDMD: AI-Powered Digital Drug Music Detector}
%
%
\author{Mohamed Gharzouli}
\authorrunning{M. Gharzouli}
%
\institute{Faculty of NTIC, Abdelhamid Mehri Constantine 2 University, Constantine, Algeria\\
\email{mohamed.gharzouli@univ-constantine2.dz}\\}
\maketitle              
\begin{abstract}
We present the first version of DDMD (Digital Drug Music Detector), a binary classifier that distinguishes digital drug music from normal music. In the literature, digital drug music is primarily explored regarding its psychological, neurological, or social impact. However, despite numerous studies on using machine learning in Music Information Retrieval (MIR), including music genre classification, digital drug music has not been considered in this field. In this study, we initially collected a dataset of 3,176 audio files divided into two classes (1,676 digital drugs and 1,500 non-digital drugs). We extracted machine learning features, including MFCCs, chroma, spectral contrast, and frequency analysis metrics (mean and standard deviation of detected frequencies). Using a Random Forest classifier, we achieved an accuracy of 93\%. Finally, we developed a web application to deploy the model, enabling end users to detect digital drug music.

\keywords {Digital Drug Music   \and Music classification \and Machine learning}
\end{abstract}
\section{Introduction}
Audio analysis and processing encompass a range of methodologies and techniques used to extract meaningful information from audio signals. These processes involve converting raw audio data into a structured form that can be analyzed to uncover patterns, recognize speech, categorize sounds, and perform other tasks. The advent of machine learning has significantly accelerated this field, enabling the automation of complex tasks such as audio classification, source separation, and speech recognition.
One of the most widely used applications in this domain is audio classification, which involves learning to categorize sounds and determine their corresponding class. This challenge is relevant to various real-world scenarios, such as identifying the genre of a song from music clips ~\cite{Ketan}. Typically, these applications rely on extracting relevant audio features such as melody, harmony, rhythm, and timbre from music tracks to achieve accurate classification.\\
The integration of machine learning techniques has revolutionized music classification, enabling the efficient analysis of large-scale audio datasets. These approaches improve the identification of specific music genres and tracks, a capability crucial for music streaming platforms and marketers aiming to enhance user experiences ~\cite{Barratt}, ~\cite{Hivo}, ~\cite{Martinez}.\\
However, despite extensive research and numerous applications addressing music classification, the detection of digital drug music remains largely unexplored. Digital drug music, also known as binaural beats or i-dosing, consists of audio tracks designed to produce specific psychological and physiological effects by playing different frequencies in each ear. It is believed that this music may alter brainwave function, potentially inducing changes in mental states. Digital drugs have recently garnered significant attention in popular culture, sparking debates around personal well-being and evoking discussions reminiscent of those concerning illicit substances and their impact on health ~\cite{Hesmondhalgh}, ~\cite{Barratt}. Despite the growing interest, particularly regarding the sociological, psychophysiological, and neurological implications of digital drug music, little has been done to examine this genre from a musical classification perspective.
In this context, proposing a machine learning-based solution for detecting and classifying digital drug music has become increasingly necessary for individuals, professionals, and music distributors.
\\Nonetheless, while machine learning offers novel solutions for music analysis, it also presents significant challenges. Issues related to data quality, accessibility, and the inherently subjective nature of music perception may hinder the performance of classification algorithms. Furthermore, the cultural diversity in music requires that models be trained inclusively, ensuring that varied musical expressions are adequately represented ~\cite{Cances}. As this field continues to evolve, ongoing debates about the ethical implications of digital drug music detection remain paramount, reflecting broader societal concerns about the intersection of technology and mental health ~\cite{Laskaris}.\\
This paper introduces DDMD: Digital Drug Music Detector, a machine learning-based system designed to automatically classify and identify digital drug music from conventional audio tracks. The system utilizes advanced feature extraction techniques and a trained machine-learning model to distinguish digital drug music from non-digital drug music. A carefully curated dataset of audio files, labelled as either digital drug music or non-digital drug music, was collected to train and validate the model. We outline the development, implementation, and evaluation of DDMD, underscoring its potential applications in digital music regulation and auditory health monitoring.\\
The rest of the paper is structured as follows: Section 2 provides an overview of the concept of digital drug music. Section 3 reviews related work. Section 4 discusses the development of the DDMD model. Section 5 describes a web application designed to customize DDMD. Finally, Section 6 concludes the study.

\section{The Science and Psychology of Digital Drugs}
Digital drugs, also known as binaural beats, audio drugs or iDoser, refer to sound waves designed to affect brainwave activity and produce effects similar to those of psychoactive substances. They work by playing slightly different frequencies in each ear, which the brain processes as a third frequency that can allegedly influence brain waves and consciousness. Another form of digital drug music is Isochronic Tones, which are single tones that turn on and off at regular intervals. These are considered more effective because they do not require headphones ~\cite{refurl1}, ~\cite{Barratt}. \\
The auditory phenomenon of binaural beats is said to activate numerous neural pathways, thus causing sensations in the listener that are either care or euphoria, resembling the effects of chemical drugs like cannabis and MDMA. Research has thus far investigated the role of these sound waves in cognitive and emotional states ~\cite{Garcia1}, ~\cite{Garcia2}. 
The digital drugs hereby aim to synchronize the brain's electrical activity, which is the process of brainwave entrainment. The synchronized brainwaves are the source of perceptual, emotional, and even out-of-body experiences, which can be compared to the effects of psychoactive drugs ~\cite{Becher}, ~\cite{Gao}. 
Psychologically, digital drugs are quite influential as they stimulate the brain's reward system and can be addictive. The psychological attractiveness of digital drugs is immense, particularly for people looking for chemical-free ways to relax or feel better ~\cite{Chaieb1}, ~\cite{Chaieb2}. According to a study conducted on more than 30,000 people from 22 different countries ~\cite{Barratt}, it was found that 72\% of them used binaural beats mainly for relaxation or sleep, while 12\% of them were looking for the same effects as real drugs. This phenomenon is, thus, a sign of the increasing interest in digital drugs as a means of mental health care, with possible applications in therapeutic settings for anxiety, insomnia, and pain relief. Here, are some key points about digital drugs ~\cite{Barratt}:
\begin{itemize}
    \item Proponents claim that binaural beats can induce altered states, relaxation, sleep, and even mimic the effects of psychoactive drugs.
    \item	However, the effects are debated and research on their efficacy is limited and mixed
    \item	Some people, especially younger individuals, use digital drugs to relax, change their mood, or try to achieve drug-like effects.
     \item	There are concerns that digital drugs could lead to curiosity about actual drugs, but they do not contain any substances and are not physically addictive.
     \item Potential risks include headaches, insomnia, anxiety, and exacerbating conditions like epilepsy.
\end{itemize}
As a result, it can be concluded that digital drugs are gaining popularity, yet further research is needed to comprehend their mechanisms and potential medical applications or risks completely. They seem to have unique effects, and the personal experience may differ for every individual. Moderation and caution are advised, as the long-term effects of digital drugs are not yet fully understood.

\section{Related Work}
In the field of Music Information Retrieval (MIR), automatic genre classification plays a crucial role, enabling websites and music stores to organize, categorize, and recommend music content efficiently ~\cite{Correa}, ~\cite{DELDJOO}. In recent years, machine learning techniques have gained significant attention from researchers in MIR, being applied to a broad range of classification problems. Numerous studies have provided summaries of audio features and classification algorithms for music genre categorization.
Most studies in music classification focus on extracting characteristics directly from audio tracks. Initially, genre recognition relied on traditional machine learning models such as decision trees, Support Vector Machines (SVMs), Random Forests, and Artificial Neural Networks (ANNs) ~\cite{Li2024}. Commonly extracted audio features include Spectral Features (e.g., Spectral Centroid, Spectral Contrast) ~\cite{Lee}, Chroma Features ~\cite{Shi}, Spectrograms ~\cite{Nirmal}, ~\cite{Lahovnik}, and Mel-Frequency Cepstral Coefficients (MFCCs). Among these, MFCC is regarded as one of the most effective techniques for audio signal processing due to its high resolution, enabling the detection of even minor changes in the signal ~\cite{Li2024}.
Despite the extensive body of work on music genre classification, no significant research has addressed the detection and classification of digital drug music. The only exception is the work of ~\cite{Abdulaziz}, which proposed a method to extract key features of digital drugs using contourlet transformation. The study aimed to identify the main characteristics of digital drug music by analyzing music signals, focusing on features such as Energy, Correlation, and Homogeneity. However, the study presents a feature extraction method without integrating a classification model. Additionally, it emphasizes high-frequency coefficients derived from contourlet transformation, potentially overlooking other crucial features for classifying digital drug music. Moreover, the effectiveness of this technique may be affected by the limited scope of the study, which tested the method on only two files, and no specific dataset was collected or employed. This limitation raises concerns about the generalizability of the findings of this study.\\
While few machine-learning-based approaches have been proposed for the classification of digital drug music, several recent studies have begun to explore it from sociological, psychological, and neurological perspectives ~\cite{Barratt}, ~\cite{Trimble}, ~\cite{Sujeesha}, ~\cite{Shao}, ~\cite{LeeMW}. As research in this area advances, exploring the underlying mechanisms and potential applications of digital drugs in healthcare and well-being will become increasingly important. Therefore, collecting a dedicated dataset and developing an AI-powered tool for digital drug music detection and classification could significantly contribute to multiple research domains related to this topic.

\section{DDMD development}
To achieve our objective, we followed the process outlined in Figure 1, which consists of two main phases. The first phase is performed manually, while the second phase encompasses the key steps in creating the end-to-end model.
\begin{figure}
\centering
\includegraphics[width=0.7\textwidth, height=0.7\textheight]{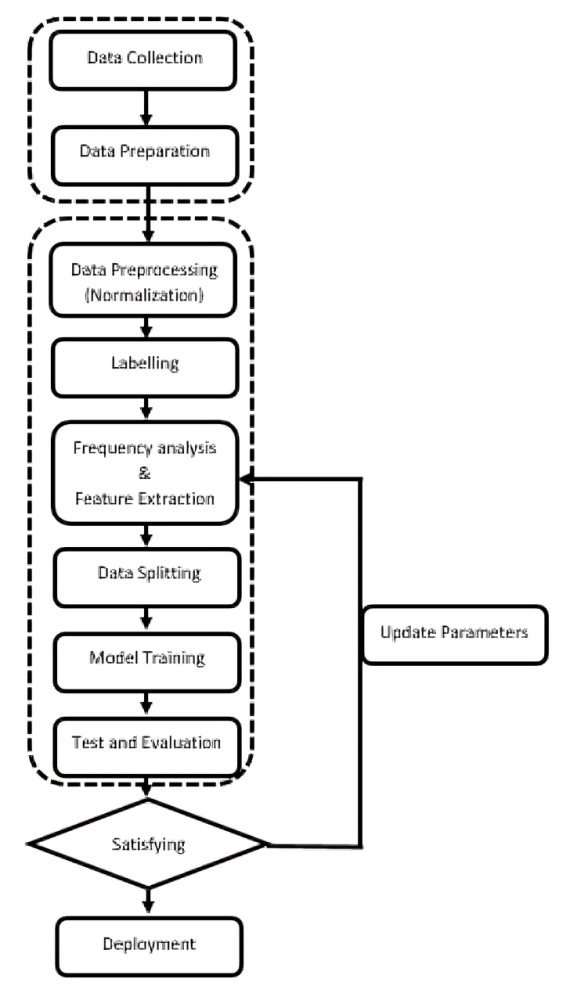}
\caption{Development Process} \label{fig1}
\end{figure}

\subsection{Data collection }
Various AI audio datasets are available, encompassing speech, music, and sound effects, and can be utilized as training data for generative AI, AI model development, intelligent audio tools, and audio applications ~\cite{refurl2}. For music analysis and processing, the field covers a range of subjects from low-level audio features to high-level musical concepts, including uses in Music Information Retrieval (MIR), audio editing, and computational musicology, enabling research in areas like music transcription, source separation, classification, and audio synthesis. With these covered topics, there are several collections of datasets like Musicnet ~\cite{musicnet1}, ~\cite{musicnet2}, Million Song Dataset ~\cite{ismir2011}, MedleyDB 2.0 ~\cite{MedleyDB}, SALAMI ~\cite{SALAMI} and others.\\
Despite the availability of various datasets, including those intended for music genre classification like GTZAN  ~\cite{GTZAN} and Extended-Ballroom  ~\cite{Ballroom}, none are specifically dedicated to digital drug music detection or classification. To address this gap, we created our new dataset, carefully considering ethical and copyright issues by selecting primarily non-copyrighted and free-licensed tracks or by extracting short segments from longer audio files.\\
The dataset consists of 3,176 MP3 files of varying quality (MP3 format was used due to storage limitations). These files were carefully and manually divided into two classes: Digital Drug (DD), containing 1,676 files, and Non-Digital Drug (NDD), comprising 1,500 files. The NDD class includes music from various genres, excluding digital drug music.
The dataset is available at the following link:\\ \url{https://www.kaggle.com/datasets/mohamedgharzouli/digital-drug-music/data}.

\subsection{Model development}
After importing the necessary libraries, particularly those used for audio processing and peak detection in signals, and following the upload and labeling of data (DD and NDD), the model development process consists of the following steps:
\subsubsection{Preprocessing}

The preprocessing step consists of the normalization of the audio data. Audio normalization is the process of changing the amplitude of an audio signal to a desired range, normally between [-1,1]. This guarantees that the audio signal's amplitude is within the maximum and minimum values supported by the digital audio format. There are two main types of audio normalization: peak normalization, which adjusts the gain of the audio signal based on the highest peak amplitude, and RMS (Root Mean Square) normalization, which adjusts the gain based on the RMS value of the audio signal, which represents the average power of the signal. In our work, we have used peak normalization. It is done by dividing the signal by the maximum absolute value in the signal:
\begin{equation}
X_{\text{normalized}} = \frac{X}{\max(|X|)}
\end{equation}

\subsubsection{Frequency Analysis}
This step performs frequency analysis on the audio signal using the Fast Fourier Transform (FFT). Here's how it works:
\begin{itemize}
    \item FFT: Converts the time-domain signal data into the frequency domain. \text{fft\_data} = \text{FFT}(\text{data})
    The input 'data' is an array representing the audio signal. This signal is a series of amplitude values sampled at regular intervals over time. The Discrete Fourier Transform (DFT) converts the time-domain signal into a sequence of complex numbers that represent the amplitude and phase of each frequency component present in the signal. The FFT of a time-domain signal 'x[n]' is given by:\\
   
\begin{equation}
X[k] = \sum_{n=0}^{N-1} x[n] \cdot e^{-i \cdot 2\pi \cdot k \cdot n / N}
\end{equation}
where:\\

X[k]: is the output of the FFT at the \(k\)-th frequency bin.\\
x[n]: is the \(n\)-th sample of the input signal.\\
N: is the number of samples in the input signal.\\
\hfill $e^{-i \cdot 2\pi \cdot k \cdot n / N}$: is the complex exponential (Euler's formula), representing the complex plane's rotation.

The output of the FFT is an array of complex numbers, where each element corresponds to a frequency component of the original signal. The magnitude of each element represents the amplitude of the respective frequency, while the angle (phase) indicates the phase shift.
Where: 
\begin{itemize}
    \item Magnitude: \( |X[k]| = \sqrt{\Re(X[k])^2 + \Im(X[k])^2} \)
    \item Phase: \( \phi[k] = \text{atan2}(\Im(X[k]), \Re(X[k])) \)
\end{itemize}

\item Corresponding frequencies are calculated by:
\begin{equation}
\text{frequencies} = \frac{k}{N \cdot \Delta t}
\end{equation}
for k=0,1,2,…, N-1 where N is the number of samples, and \[
\Delta t = \frac{1}{\text{sample\_rate}}
\]
\item Peak Detection: Peaks in the magnitude spectrum are identified using a threshold (0.1 times the maximum magnitude). It involves finding frequency components where M(f) has local maxima:\\
\[
\text{If } M(f_k) > M(f_{k-1}) \text{ and } M(f_k) > M(f_{k+1}), \text{ then } f_k \text{ is a peak.} \]\\
Additionally, ensure that the magnitude \( M(f_k) \) is greater than 10\% of the maximum value:
\[
M(f_k) \geq 0.1 \times \max(M(f))
\]
For each peak index \( k_{\text{peak}} \), calculate the corresponding frequency \( f_{\text{detected}} \) using:

\begin{equation}
f_{\text{detected}} = \frac{k_{\text{peak}} \cdot f_s}{N}
\end{equation}
Where: \( f_s \) is the sampling rate.

\end{itemize}
In summary, the objective of the frequency analysis is to extract two Frequency Domain Features: the mean of detected frequencies and the standard deviation of detected frequencies. The mean provides a central measure of the dominant frequencies, while the standard deviation provides information about how much the frequencies deviate from the mean.\\
The mean (average) is mathematically represented as:

\begin{equation}
\text{Mean} = \mu = \frac{1}{n} \sum_{i=1}^{n} f_i
\end{equation}
where:
\begin{itemize}
    \item \( n \) is the number of detected frequencies,
    \item \( f_i \) are the values of the detected frequencies.
\end{itemize}
The formula for standard deviation is:
\begin{equation}
\text{Standard Deviation} = \sigma = \sqrt{\frac{1}{n} \sum_{i=1}^{n} (f_i - \mu)^2}
\end{equation}
Where:
\begin{itemize}
    \item \( \mu \) is the mean of the detected frequencies (computed above),
    \item \( f_i \) are the detected frequencies,
    \item \( n \) is the number of detected frequencies.
\end{itemize}
At the end of this step, all frequency analysis features are calculated and concatenated into a single feature vector.

\subsubsection{Feature Extraction}
In this step, we extract 32 machine-learning features from the audio signal:
\begin{itemize}
    \item \textbf{MFCCs}: In our case, we opted to use 13 MFCCs. This decision aligns with the nature of our collected dataset and our goal to develop an efficient, simple model. The number of MFCCs is chosen based on factors such as the specific application, computational efficiency, and the characteristics of the audio data. While a higher number of MFCCs captures more spectral detail, which may be necessary for complex audio tasks, using fewer MFCCs reduces the dimensionality of the feature set, simplifying the model and speeding up computation. This helps avoid increased computational load and the risk of overfitting.\\
    Especially when the dataset consists of tracks with consistent musical content, such as isochronic tones, fewer MFCCs may be sufficient due to less variation to capture. Additionally, the duration of the audio tracks influences the choice of MFCCs. Shorter tracks provide less data, so fewer MFCCs may suffice to capture essential characteristics. Using too many MFCCs on short tracks risks overfitting, as the model might capture noise or irrelevant details instead of the core signal features.
    \item  \textbf{Chroma Features}: the number of coefficients is usually set to 12, corresponding to the 12 different pitch classes (or chroma) in the Western music notation  ~\cite{korea}.
    \item \textbf{Spectral Contrast}: Measure the difference between peaks and valleys in a sound spectrum. In our work as an initial experiment, we used 7 spectral contrast coefficients. This range is generally sufficient to capture the necessary detail in the frequency distribution across different bands without overwhelming the model with too much information.
\end{itemize}
 Finally, we combined the features extracted from frequency analysis (mean and standard deviation of detected frequencies) with the machine learning features (MFCCs, chroma, and spectral contrast). This resulted in a total of 34 extracted features.
\subsubsection{Model creation}
When choosing a classifier for audio classification tasks, several factors should be considered, including the nature of the data, the computational resources available, and the desired performance characteristics. In our case, as a first choice, we have used Random Forest. This classifier is relatively easy to implement and tune. It requires fewer hyperparameters to be set, and default settings often perform well. In addition, it is robust to noise and irrelevant features and one of its key advantages is its ability to provide insights into feature importance ~\cite{Pavan},  ~\cite{Master}. Furthermore, Random Forest can be parallelized easily, for this reason, we create our model by processing a batch of audio files in parallel using the \textit{ProcessPoolExecutor}. 
To create our model, we followed these steps:
\begin{itemize}
    \item \textbf{Data Splitting}: The dataset is split into training and testing sets (80\% train, 20\% test). 
     \item \textbf{Training}: The Random Forest Classifier is trained using the training data.
     \item \textbf{Model Evaluation}: The model's accuracy is calculated, and a classification report is generated.
\end{itemize} 
The obtained results were very encouraging with a test accuracy of 93\%. The results are summarized in table 1:
\begin{table}
\centering
\caption{Classification Report.}\label{tab1}
\begin{tabular}{|l|c|c|c|c|}
\hline
& \textbf{Precision} & \textbf{Recall} & \textbf{F1-Score} & \textbf{Support} \\
\hline
\textbf{Class 0} & 0.91 & 0.94 & 0.92 & 280 \\
\textbf{Class 1} & 0.95 & 0.93 & 0.94 & 355 \\
\hline
\textbf{Accuracy} & \multicolumn{3}{c|}{0.93} & 635 \\
\textbf{Macro avg} & 0.93 & 0.93 & 0.93 & 635 \\
\textbf{Weighted avg} & 0.93 & 0.93 & 0.93 & 635 \\
\hline
\end{tabular}
\end{table}

\section{Customization and deployment}
The work presented in the previous section presents the first phase of the DDMD project. To leverage the first obtained results of this phase, we initially thought about developing a web application to customize the first version of the DDMD model. This web application classifies audio files as either Digital Drug music (DD) or Non-Digital Drug music (NDD). Users can upload an audio file or provide a YouTube URL, and the application will process the input, extract features, and utilise the pre-trained Random Forest model to make predictions.
The web application provides many features like:
\begin{itemize}
    \item Accepts various audio file formats: WAV, MP3, MP4, AIFF, AAC, OGG Vorbis, and FLAC.
\item Converts non-MP3 files to MP3 format for consistent processing.
\item Allows input via a YouTube URL, extracting the audio and converting it to MP3.
\item Validates the file type and enforces a maximum file size of 50MB (can be modified by the developer).
\item Displays the classification result (DD or NDD).
\item Handles possible errors and exceptions.
\end{itemize}
Figure 2 presents the user interface of this web application. It is developed with Flask. It is located in the sub-folder 'DDMCWebAppé of DDMD project.   
\begin{figure}
\includegraphics[width=\textwidth]{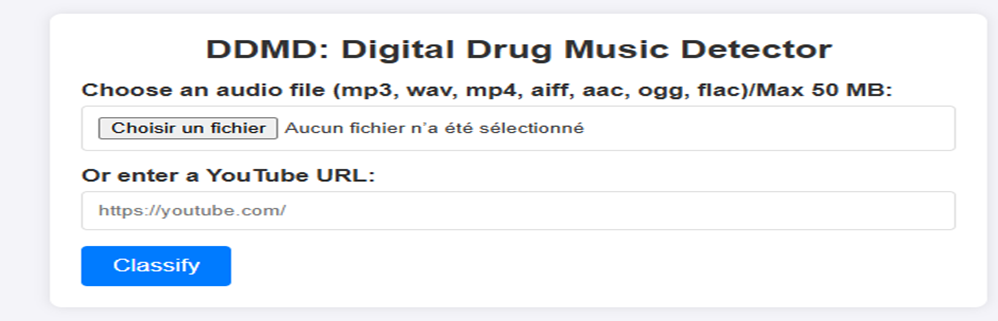}
\caption{User interface of the web application} \label{fig2}
\end{figure}
The web application with the random forest model and the obtained results are available at the following link:\\
\url{https://www.kaggle.com/models/mohamedgharzouli/ddmd-digital-drug-music-detector}.
The sub-folder \texttt{DDMD\_WebAPP} contains the Flask-based web project, including all details about the web application's technologies, prerequisites, installation, and usage.

\section{Conclusion and future work}
In recent years, digital drug music has gained significant popularity. Numerous studies have been conducted to explore the relationship between binaural beats and brain activity, and to uncover the effects of this type of music on the brain. These research efforts have primarily focused on the sociological, psychological, and neurological influences of the digital drug phenomenon. However, despite the advancements in music categorization, the classification or detection of digital drug music using machine learning techniques has been largely overlooked. Developing an intelligent classification tool for digital drug music could be highly beneficial and may complement other studies, further advancing research in this area.\\
In this paper, we have presented DDMD (Digital drug music detector) a machine learning-based tool that distinguishes digital drug and non-digital drug music. The development of DDMD began with the collection of a large dataset comprising 1,676 digital drug music tracks and 1,500 tracks of conventional music from various genres. Then, we developed a Random Forest classifier that uses many extracted features. We combined the features extracted from frequency analysis (mean and standard deviation of detected frequencies) with the machine learning features (MFCCs, chroma, and spectral contrast).\\
The initial results are promising, achieving an accuracy of 93\%, which demonstrates the effectiveness of a machine learning approach in detecting digital drug music. However, it is important to note that this paper presents the first version of DDMD, which has some notable limitations:

\begin{itemize}
    \item Conduct more in-depth experiments to empirically determine the optimal number of MFCCs. By testing different numbers of coefficients and evaluating model performance, we can identify the optimal number that balances accuracy and computational efficiency for this specific application.
     \item It is essential to compare the performance of the Random Forest model with other models, such as SVMs or neural networks, using cross-validation and other evaluation methods.
\end{itemize}
Moreover, this study could be further enhanced through the following steps:
\begin{itemize}
    \item Utilize the extracted features to improve detection accuracy through feature-based transfer learning. We are currently developing new models. An example of a simple student-teacher model is available at: \url{https://www.kaggle.com/code/mohamedgharzouli/ddmd-student-teacher-model}.
    \item  Develop the second version of DDMD, referred to as DDMC (Digital Drug Music Classifier), which will be a multi-class classifier capable of distinguishing between various types of digital drug music (e.g., deep sleep, relaxation and meditation, active thinking, alertness, etc.).
\end{itemize}

\end{document}